\documentclass[12pt]{article}

\begin{document}

\title{ The Black-Scholes Equation and Certain Quantum
Hamiltonians  }

\author{ Juan M. Romero$^{1}$\thanks{jromero@correo.cua.uam.mx},
O. Gonz\'alez-Gaxiola$^{2}$\thanks{ogonzalez@correo.cua.uam.mx},
J. Ru\'iz de Ch\'avez$^{3}$\thanks{jrch@xanum.uam.mx},\\
R. Bernal-Jaquez$^{4}$\thanks{rbernal@correo.cua.uam.mx},\\
\it $^{1,2,4}$ Departamento de Matem\'aticas Aplicadas y Sistemas,\\
\it Universidad Aut\'onoma Metropolitana-Cuajimalpa\\
\it M\'exico, D.F  01120, M\'exico\\
\it $^{3}$ Departamento de Matem\'aticas,\\
\it Universidad Aut\'onoma Metropolitana-Iztapalapa\\
\it Av. San Rafael Atlixco No. 186, Col. Vicentina, A.P. 55--534,\\
\it  09340 Izta\-palapa,
 M\'exico, D.F.}

\date{}

\pagestyle{plain}

\maketitle

\begin{abstract} In this paper a quantum mechanics is
built by means of a non-Hermitian momentum operator. We have shown
that it is possible to construct two Hermitian and two non-Hermitian
type of Hamiltonians using this momentum operator. We can construct
a generalized supersymmetric quantum mechanics that has a dual based
on these Hamiltonians. In addition, it is shown that the
non-Hermitian Hamiltonians of this theory can be related to
Hamiltonians that naturally arise in the so-called quantum finance.

\end{abstract}



\section{Introduction}
\label{s:Intro}
In the origins of quantum mechanics P.A.M  Dirac \cite{Dirac:gnus}
observed that commutation relations
\begin{eqnarray}
[x_{i},x_{j}]=[P_{i},P_{j}]=0,\qquad [x_{i},P_{j}]=i \delta_{ij},
\end{eqnarray}
are satisfied by the operators $x_{i}, P_{j}= -i\partial_{j}$ and
also by the set
\begin{eqnarray}
x_{i},\qquad P_{(f)j}= -i\partial_{j} +i\partial_{j}f,
\label{eq:no-herm}
\end{eqnarray}
with $f$ been  an arbitrary function \cite{Dirac:gnus}. For
different reasons, operators (\ref{eq:no-herm}) were discarded, for
example, the Hamiltonian  $H_{f}=\frac{P_{f}^{2}}{2}$ is
non-Hermitian and it could have a non-real spectrum. However, it has
recently been shown that there are non-Hermitian operators with real
spectrum \cite{bender-0:gnus}. Studies of non-Hermitian Hamiltonians
and their applications in physics
can be found in the papers \cite{bender-1:gnus}, \cite{bender-2:gnus} and \cite{bender-3:gnus} .\\

In this paper, we will show that, using the operator $P_{(f)j}$,
four different types of  Hamiltonians can be built, two of these
Hermitians and the other two, non-Hermitians.  We will show that,
from two Hermitian Hamiltonians in one dimension, it is possible to
construct a supersymmetric mechanics, and that  using one of the two
non-Hermitian Hamiltonians a generalized supersymmetric mechanic can
be constructed. Moreover, we will  show that this new supersymmetric
quantum mechanics
has a dual and the ground state of the corresponding Hamiltonians will be found.\\

As a second  point  of this work, it is shown how  the operator
$P_{(f)j}$ can
also  be used to build some of non-Hermitian Hamiltonians that naturally  arise in the so-called quantum finance.\\

\section{Non-Hermitian Hamiltonians}

In this section, we will study the quantum mechanics that emerges
when the operator $P_{(f)j}$ is considered. As an starting point, we
have to notice that the operator  $P_{(f)i} $ is given by the
transformation
\begin{eqnarray}
P_{(f)i} = e^{f}P_{i}e^{-f}, \label{eq:momentum}
\end{eqnarray}
and also  that it is  not Hermitian.\\

With $\vec P_{(f)}$ we can construct four Hamiltonians, two of them
Hermitians
\begin{eqnarray}
H_{1}&=&\alpha^{2} \vec P_{(f)}^{\dagger} \cdot \vec P_{(f)}=\alpha\left(\vec P^{\;2}+\nabla ^{2} f+\left(\vec \nabla f \right)^{2}\right), \\
H_{2}&=& \alpha^{2} \vec P_{(f)} \cdot \vec
P^{\dagger}_{(f)}=\alpha\left(\vec P^{\;2}-\nabla ^{2} f+\left(\vec
\nabla f \right)^{2}\right)
\end{eqnarray}
and two non-Hermitians
\begin{eqnarray}
H_{3}&=&\beta^{2} \vec P_{(f)}^{\dagger}\cdot \vec P_{(f)}^{\dagger}\nonumber\\
&=&\beta^{2} \left(\vec P^{\;2}-2i\vec \nabla f \cdot \vec P -
\nabla ^{2} f -
\left(\vec \nabla f \right)^{2} \right),\\
H_{4}&=& \beta^{2} \vec P_{(f)}\cdot \vec P_{(f)}\nonumber \\
&=&\beta^{2}\left(\vec P^{\;2}+2i\vec \nabla f \cdot \vec P + \nabla
^{2} f - \left(\vec \nabla f \right)^{2} \right).
\end{eqnarray}
These Hamiltonians are obtained naturally in different contexts. In
the following subsection, we will see that, they can be used to
obtain a generalized version of the supersymmetric quantum
mechanics.

\section{Supersymmetric Quantum Mechanics }

In the one dimensional  case, the Hamiltonians $H_{1}$ and $H_{2}$
are given by
\begin{eqnarray}
H_{1}&=&\alpha^{2}\left( P^{\;2}+\frac{d^{2}f}{dx^{2}} +\left( \frac{df}{dx} \right)^{2} \right), \\
H_{2}&=& \alpha^{2}\left( P^{\;2} -\frac{d^{2}f}{dx^{2}}+ \left(
\frac{df}{dx} \right)^{2}  \right).
\end{eqnarray}
Moreover, if
\begin{eqnarray}
f(x)=\int_{0}^{x}W(u)du,
\end{eqnarray}
then
\begin{eqnarray}
H_{1}&=&\alpha^{2}\left( P^{\;2}+\frac{dW}{dx} +W^{2} \right), \\
H_{2}&=& \alpha^{2}\left(P^{\;2}-\frac{dW}{dx}+ W^{2}  \right).\\
\end{eqnarray}
This Hamiltonians can be used to form the matrix
\begin{eqnarray}
h&=&\left(
\begin{array}{rr}
 H_{1} & 0\\
 0&  H_{2}
\end{array}
\right).
\end{eqnarray}
Now, defining
\begin{eqnarray}
Q= \left(
\begin{array}{rr}
 0& \alpha  P_{(f)}\\
 0& 0
\end{array}
\right),
\end{eqnarray}
we have
\begin{eqnarray}
h=\{Q,Q^{\dagger}\},\qquad Q^{2}=0,\quad \{Q,H\}=0.
\end{eqnarray}
According to the supersymmetric quantum mechanics
\cite{cooper:gnus}, $h$  represents a superhamiltonian
 and $Q$ a supercharge. Therefore, the quantum mechanics  built using
 $P_{(f)i}$ contains the usual supersymmetric quantum mechanics.\\

Moreover, $P_{(f)}$ allows us to generalize supersymmetric quantum
mechanics. In fact, we can define the matrices
\begin{eqnarray}
\quad Q_{1}= \left(
\begin{array}{rrrr}
 0& \alpha  P_{(f)}& 0 & 0\\
 0& 0         & 0 & 0        \\
0 & 0 & 0& \beta  P_{(f)}^{\dagger}\\
0 &0 & 0&0
\end{array}
\right),\, Q_{2}= \left(
\begin{array}{rrrr}
 0& 0 & 0 & 0\\
\alpha  P_{(f)}^{\dagger} & 0         & 0 & 0        \\
0 & 0 & 0&  0\\
0&0 & \beta P_{(f)}^{\dagger} &0
\end{array}
\right),
\end{eqnarray}
and then  $Q_{1}^{2}=Q_{2}^{2}=0$. Using  $Q_{1}^{2}=Q_{2}^{2}=0$ we
can construct the Hamiltonian
\begin{eqnarray}
H&=& \{Q_{1},Q_{2}\}=\left(
\begin{array}{rrrr}
 H_{1} & 0& 0 & 0\\
 0&  H_{2}& 0 & 0        \\
0 & 0 &  H_{3}&  0\\
0 &0 & 0&  H_{3}
\end{array}
\right)
\end{eqnarray}
with the conserved charges
\begin{eqnarray}
\dot Q_{1}=\left[Q_{1},H\right]=0,\qquad \dot
Q_{2}=\left[Q_{2},H\right]=0;
\end{eqnarray}
now, if $\beta=0,$ we have $Q_{1}=Q_{2}=Q$ and this quantum mechanics reduces to the
usual supersymmetric quantum mechanics.\\

Besides, we have another quantum mechanics that reduces to the usual
supersymmetric quantum mechanics. In fact, if
\begin{eqnarray}
\quad Q_{3}= \left(
\begin{array}{rrrr}
 0& \alpha  P_{(f)}^{\dagger}& 0 & 0\\
 0& 0         & 0 & 0        \\
0 & 0 & 0& \beta  P_{(f)}\\
0 &0 & 0&0
\end{array}
\right),\, Q_{4}= \left(
\begin{array}{rrrr}
 0& 0 & 0 & 0\\
\alpha  P_{(f)}& 0         & 0 & 0        \\
0 & 0 & 0&  0\\
0&0 & \beta  P_{(f)} &0
\end{array}
\right),
\end{eqnarray}
then  $Q_{3}^{2}=Q_{4}^{2}=0$ and  we can construct the Hamiltonian
\begin{eqnarray}
\tilde H&=&\{Q_{3},Q_{4}\}= \left(
\begin{array}{rrrr}
 H_{2} & 0& 0 & 0\\
 0&  H_{1}& 0 & 0        \\
0 & 0 &  H_{4}&  0\\
0 &0 & 0&  H_{4}
\end{array}
\right).
\end{eqnarray}
Note that if $\beta=0$ this Hamiltonian is just the usual superhamiltonian $h.$\\

 If we make the transformation $f\to -f,$  we have
\begin{eqnarray}
\left(H_{1}, H_{2}, H_{3}, H_{4}\right) \qquad \to \qquad
\left(H_{2}, H_{1}, H_{4}, H_{3}\right),
\end{eqnarray}
{\it i.e}
\begin{eqnarray}
H\to \tilde H.
\end{eqnarray}
Then, there is a duality transformation between Hamiltonians $H$ and
$\tilde H.$
Therefore these generalized quantum mechanics are duals. \\

Now, if we consider the  functions  $\psi_{0}=A_{1}
e^{f},\phi_{0}=A_{2}e^{-f}$ that satisfy
\begin{eqnarray}
 P_{(f)}\psi_{0}=0, \qquad  P_{(f)}^{\dagger}\phi_{0}=0.
\end{eqnarray}
Then,  the wave function
\begin{eqnarray}
\psi&=&\left(
\begin{array}{r}
A_{1} e^{-f} \\
A_{2} e^{f}\\
A_{3} e^{-f}\\
A_{3} e^{-f}
\end{array}
\right)
\end{eqnarray}
satisfies
\begin{eqnarray}
H\psi=0.
\end{eqnarray}
Moreover, the wave function
\begin{eqnarray}
\tilde \psi&=&\left(
\begin{array}{r}
 e^{f} \\
 e^{-f}\\
 e^{f}\\
 e^{f}
\end{array}
\right)
\end{eqnarray}
satisfies
\begin{eqnarray}
\tilde H\tilde \psi=0.
\end{eqnarray}
Thus, $\psi$ is the ground state of $H$ and $\tilde \psi$ is the ground state of $\tilde H.$\\

\section{ The Black-Scholes model}
This model is a partial differential equation whose solution
describes the value of an European Option. See
\cite{Black-Scholes}, \cite{Merton}.  Nowadays, it is widely used to
estimate the pricing of options other than the European ones. Let
$(\Omega, \mathcal{F}, P,\mathcal{F}_{t\geq 0}) $ be a filtered
probability space and let $W_t$ be a brownian motion in
$\mathbf{R}$. We will consider the stochastic differential equation
(s.d.e.)
\begin{equation}
dX(t) = a(t,X(t))dt + \sigma(t, X(t))dW(t),\nonumber
\end{equation}
with $a$ and $\sigma$ continuous in $(t,x)$ and Lipschitz  in $x$.
The price processes  given by the geometric brownian  motion
$S(t),\,  S(0)= x_0$, solution of the s.d.e.
\begin{equation}
dS(t) = \mu S(t)dt + \sigma S(t)dW(t),\nonumber
\end{equation}
with $\mu $ and $\sigma$ constants. It is well know the solution of
this s.d.e. it is given by:
\begin{equation}
dS(t) = x_0\exp\{ \sigma(W(t)-W(t_0))
+(r-\frac{1}{2}\sigma^2)(t-t_0)\} \nonumber
\end{equation}
Let $0\leq t< T$  and $h$ be a Borel measurable function, $h(X(T))$
denote the contingent claim, let $\displaystyle E^{x,t}h(X(T))$ be
the expectation of $h(X(T))$, with the initial condition  $X(t)= x$.

Now we recall the Feynman--Kac theorem \cite{shreve2}. Let
$\displaystyle v(t,x)= E^{x,t}h(X(T))$ be, $0\leq t < T$, where
$dX(t) = a(X(t))dt + \sigma(X(t))dW(t)$. Then
\begin{eqnarray}\label{FK}
v_t(t,x)+ a(x)v_x(t,x)+\frac{1}{2}\sigma^2(x)v_{xx}(t,x)= 0, \,
\rm{and } \quad v(T,x) = h(x).
\end{eqnarray}
Now, if we consider the discounted value
$$u(t,x)=
e^{-r(T-t)}E^{x,t} h(X(T))=e^{-r(T-t)}v(t,x).$$ Then if at time $t$,
$S(t)= x$, if we  proceed in standard way,
\begin{eqnarray}
v(t,x)&=& e^{r(T-t)}u(t,x), \nonumber\\
 v_t(t,x)&=& -re^{r(T-t)}u(t,x)
+e^{r(T-t)}u_t(t,x), \nonumber\\
 v_x(t,x)&=& e^{r(T-t)}u_x(t,x), \nonumber\\
v_{xx}(t,x) &=& e^{r(T-t)}u_{xx}(t,x).\nonumber
\end{eqnarray}
The Black-Scholes equation is obtained substituting the above
equalities in the equation (\ref{FK}) and multiplying by the
factor $e^{-r(T-t)}$:
\begin{eqnarray} \label{BS}
& -ru(t,x) +  u_t(t,x) +  rxu_x(t,x) +
\frac{1}{2}\sigma^2x^2v_{xx}(t,x)= 0,& \\
& 0\leq t< T, x\geq 0.&
 \nonumber
\end{eqnarray}

\section{The Relation with the Black-Scholes Equation}

The operators $\vec P_{f}\cdot \vec P_{f}$ and $\vec
P_{f}^{\dagger}\cdot \vec P_{f}^{\dagger}$ are non-Hermitians, using
them only non-Hermitians Hamiltonians  such as $H_{3}$ and $H_{4},$
can be constructed. However, we will show that these operators may
have applications in some other areas  such as quantum finance. In
order to see this, we define the potentials
\begin{eqnarray}
U_{1}(x,y,z)&=&-\beta^{2}\left( \nabla ^{2} f -
\left(\vec \nabla f \right)^{2} \right)+V_{1}(x,y,z),\\
U_{2}(x,y,z)&=& \beta^{2}\left( \nabla ^{2} f +\left(\vec \nabla f
\right)^{2} \right)+V_{2}(x,y,z),
\end{eqnarray}
and the non-Hermitians Hamiltonians
\begin{eqnarray}
H_{I}&=& \beta^{2} \vec P_{(f)} \cdot \vec P_{(f)}+U_{1}(x,y,z) \nonumber\\
&=&\beta^{2}\left(\vec P^{\;2}+2i\vec \nabla f \cdot \vec P \right)+ V_{1}(x,y,z),\\
H_{II}&=&\beta^{2} \vec P_{(f)}^{\dagger}\cdot \vec P^{\dagger}_{(f)}+U_{2}(x,y,z)\nonumber\\
&=&\beta^{2}\left(\vec P^{\;2}-2i\vec \nabla f \cdot \vec P \right)+
V_{2}(x,y,z).\label{eq:pt-bs}
\end{eqnarray}

On the other hand, let us consider  the fundamental equation in
quantum finance, the so-called Black-Scholes equation (\ref{BS})
\begin{eqnarray}
\frac{\partial C }{\partial t}= - \frac{\sigma^{2} S^{2}}{2}
\frac{\partial ^{2} C} {\partial S^{2}} -rS \frac{\partial
C}{\partial S}+rC,
\end{eqnarray}
where $C$ is the option price, $\sigma$ is a constant called the
volatility and $r$ is the interest rate \cite{baaquie:gnus}. With
the change of variable $S=e^{x}$ we obtain
\begin{eqnarray}
\frac{\partial C }{\partial t}&=& H_{BS}C,\nonumber\\
H_{BS}&=&- \frac{\sigma^{2}}{2} \frac{\partial ^{2} } {\partial
x^{2}}+ \left(   \frac{\sigma^{2}}{2} -r \right)\frac{\partial }
{\partial x}+r
\end{eqnarray}
this non-Hermitian Hamiltonian is called Black-Scholes Hamiltonian.
Now, considering the one dimensional  case of  (\ref{eq:pt-bs}) and
identifying
\begin{eqnarray}
\beta^{2}=\frac{\sigma^{2}}{2},\quad
f(x)=\frac{1}{\sigma^{2}}\left(\frac{\sigma^{2}}{2}-r\right)x,\quad
V_{2}(x)=r\nonumber
\end{eqnarray}
we obtain   $H_{II}=H_{BS}.$\\

One generalized  Black-Scholes equation, (see  \cite{baaquie:gnus})
is given by
\begin{eqnarray}
H_{BSG}&=&- \frac{\sigma^{2}}{2} \frac{\partial ^{2} } {\partial
x^{2}}+ \left(   \frac{\sigma^{2}}{2} -V(x) \right)\frac{\partial }
{\partial x}+V(x).
\end{eqnarray}
In this case, considering again the one dimensional  case of
equation (\ref{eq:pt-bs}) and with
\begin{eqnarray}
\beta^{2}=\frac{\sigma^{2}}{2},\quad f(x)=\int_{0}^{x}du
\frac{1}{\sigma^{2}}\left(\frac{\sigma^{2}}{2}-V(u)\right),\quad
V_{2}(x)=V(x)\nonumber
\end{eqnarray}
we have $H_{II}=H_{BSG}.$\\

Moreover, the so-called barrier option case has  Hamiltonian
\begin{eqnarray}
H_{BSB}&=&- \frac{\sigma^{2}}{2} \frac{\partial ^{2} } {\partial
x^{2}}+ \left(   \frac{\sigma^{2}}{2} -r \right)\frac{\partial }
{\partial x}+V(x).
\end{eqnarray}
Again,  considering (\ref{eq:pt-bs})  in one dimension and with
\begin{eqnarray}
\beta^{2}=\frac{\sigma^{2}}{2},\quad
f(x)=\frac{1}{\sigma^{2}}\left(\frac{\sigma^{2}}{2}-r\right)x,\quad
V_{2}(x)=V(x) \nonumber
\end{eqnarray}
we have $H_{II}=H_{BSB}.$\\

As we have seen,  several important Hamiltonians appearing in
quantum finance are particular cases of this new version of quantum
mechanics.

\section{Summary}
\label{s:Summ}
A quantum mechanics is built  by means of a non-Hermitian momentum
operator. Moreover, It is shown that using this momentum operator
 it is possible to construct  two Hermitian and two non-Hermitian type of Hamiltonians.
Using these  Hermitian Hamiltonians we have built, a generalized
supersymmetric quantum mechanics with a dual that can be
constructed. It also shown that, the non-Hermitian Hamiltonians of
this theory may be related to so-called quantum finance Hamiltonian.


\bigskip
\begin{thebibliography}{00}

\bibitem{baaquie:gnus}
B. E. Baaquie, {\it Quantum Finance,} Cambridge University Press
(2004).

\bibitem{Black-Scholes} F. Black and M. Scholes,  The Pricing Options and Corporate
Liabilities, {\it Journal of Political Economy }(81), 637-659
(1973).


\bibitem{bender-0:gnus}
C. M. Bender, Introduction to PT-Symmetric Quantum Theory, {\it
Contemp. Phys}. {\bf 46}, 277 (2005).


\bibitem{bender-2:gnus} C. M. Bender, P. D. Mannheim,  No-ghost theorem for the fourth-order derivative
Pais-Uhlenbeck oscillator model  {\it Phys. Rev. Lett.} {\bf 100},
110402  (2008).


\bibitem{cooper:gnus}
F. Cooper, A. Khare, U. Sukhatme,  Supersymmetry and Quantum
Mechanics, {\it Phys. Rept.} {\bf 251}, 267-385 (1995).

\bibitem{Dirac:gnus} P.A. M. Dirac, \textit{The Principles of Quantum Mechanics}, Oxford (1930).

\bibitem{bender-1:gnus} P. K. Ghosh,  On the construction of a pseudo-Hermitian quantum
system with a pre-determined metric in the Hilbert space, {\it  J.
Phys. {\bf A}, Math. Theor.} { \bf 43}, 125203 (2010).


\bibitem{Merton} R.C. Merton Theory of Rational Options Pricing, {\it Bell Journal of Economic and Management
Science} (4), 141-183, (1973).

\bibitem{shreve2} S. Shreve, {\it Stochastic Calculus for Finance II:
Continuous-Time Models} (Springer Finance) (2004).

\bibitem{bender-3:gnus} T. Tanaka,  General aspects of PT-symmetric and P-self-adjoint quantum theory in a Krein space,
{\it J. Phys. A} {\bf 39}, 14175  (2006).


\end{thebibliography}
\end{document}